\documentclass[%
reprint,
amsmath,amssymb,
aps,showpacs,
]{revtex4-2}
\usepackage{bm}
\usepackage{xcolor} 
\usepackage{graphicx}
\usepackage{dcolumn}
\usepackage{bm}

\begin{document}
	
	\preprint{APS/123-QED}
	
	\title{Reentrant localization transition, quantum butterfly and robust edge modes in aperiodic zig-zag ladder}
	\thanks{A footnote to the article title}%

	\author{Sayan Bhattacharya}
	\email{sayanbhatta2002@gmail.com}
	\affiliation{
		Department of Physics, Acharya Prafulla Chandra College,
		New Barrackpore, Kolkata, West Bengal-700131, India
	}
	
	\author{Rhiddha Acharjee}
	\email{rhiddhaacharjee777@gmail.com}
	\affiliation{
		Department of Physics, University of Calcutta,
		92, Acharya Prafulla Chandra Road,
		Kolkata, West Bengal-700009, India
	}
	
	\author{Atanu Nandy}
	\email{atanunandy1989@gmail.com}
	\affiliation{
		Department of Physics, Acharya Prafulla Chandra College,
		New Barrackpore, Kolkata, West Bengal-700131, India
	}

	\date{\today}
	
	\begin{abstract}
		Low dimensional tight-binding lattices in presence of quasiperiodic disorder generally exhibits localization transition. The system supports diffusive modes upto a limiting strength of disorder and all the eigenstates become localized beyond that critical strength thereby quenching the kinetic signature of the wavepacket. However, moving away from this situation, we demonstrate that with minimal long-range off-diagonal modulation, the eigenspectrum again may offer delocalization of electronic states for some subtle combination of kinetic parameters of the Hamiltonian leading to a second quantum phase change. The localization transitions are associated with the obvious presence of single-particle mobility edges. Multifractal energy landscape also shows quantum butterfly pattern with the in-gap robust edge modes. The re-emerging localization transition is manifested through the evaluation of inverse participation
		ratio, eigenspectrum and a pertinent quantum dynamical study.
		
	\end{abstract}
	\maketitle
	
	
	\section{Introduction}
	The classic demonstration of localization of single-particle eigenstate in presence of uncorrelated (random) disorder in low-dimensional quantum systems has received everlasting relevance over the past sixty years, since its initial proposition in 1958 by P. W. Anderson \cite{Anderson1958}. This pioneering work eventually generates fascinating features in quantum transport properties of randomly disordered systems. With the advancement of fabrication and lithographic mechanisms the fundamental concept has extended its realm beyond the electronic proposition and has encompassed tailor made lattices including the photonic \cite{Yablonovitch1987,John1987}, phononic \cite{Montero1998,Vasseur1998}, plasmonic \cite{Tao2007,Christ2007} or polaritonic \cite{Barinov2009,Grochol2008} lattices. Recently, ultra-cold gaseous system has even been enlisted for the direct observation of localization of matter waves \cite{Damski2003,Billy2008,Roati2008}.
    
    While the single-particle mobility edge (SPME) associated with the canonical case of Anderson localization (AL) is constrained to higher-dimensional systems, the search for a metal-insulator transition (MIT) in one dimensional tight-binding lattices has always been an intriguing field which started earlier with certain aperiodic lattices which maintain a bridge between a perfectly ordered and random system of scatterers. The remarkable evidence is the single band $\text{Aubry-Andr\'e}$  (AA) model in which the on-site energies follow a deterministic quasiperiodic fashion. This system displays a localization transition with respect to a critical strength of disorder. \cite{Aubry1980} However, another premier feature is observed in different generalized AA systems and other quasiperiodic models \cite{DasSarma1986,Biddle2010,Ganeshan2015,Sun2015,Gopalakrishnan2017,Purkayastha2017}, that the phase transformation is incidentally associated with a critical regime where both diffusive and non-conducting modes coexist. The immediate inference is the possibility of SPME that corresponds to a specific energy boundary bifurcating the extended and the insulating phases of the eigenspectrum. The notable progress in the context
of quantum gases in optical lattices, the localization transition and the existence of SPME in
quasiperiodic geometries have received considerable momentum \cite{Boers2007,Li2017} because of its experimental realizations \cite{Luschen2018,An2018,An2021}. Such systems indicating localization transition shows a monotonically enhanced localization if one goes far beyond the critical strength. But in this letter, we present a quasi-one dimensional \textit{zig-zag} ladder network in the tight-binding description which describes that with a subtle
choice of slowness index of the $\text{Aubry-Andr\'e-Harper}$ (AAH) type of next nearest neighbor hopping integral and other kinetic parameters the delocalization of electronic states may reappear in the spectrum. This simple tight-binding model with minimal off-diagonal modulation thus represents a reentrant localization transition associated with corresponding SPME. This unconventional reemerging transition is limited to exhibit in one dimensional dimerized chain of atoms \cite{Roy2021} but it is no longer reported for any two-port ladder-like geometry, to the best of our knowledge. Moreover, in most of the cases it is more likely that the strenght of the quasiperiodic disorder plays the key role in the localization transition, but in our case, we report that a slowness exponent in the AAH variation controls the phase transition hosting the SPME. It is to be noted that ladder networks play a remarkable model while considering the problem of understanding the charge transport in double stranded DNA models \cite{Macia2006,Cuniberti2007}. Mrevlishvili \cite{Mrevlishvili1998} experimentally realized oscillatory signature in the specific heat of DNA structures at low temperatures, findings which have been subsequently explained by Moreira et al. \cite{Moreira2006} taking a quasiperiodic arrangement of the nucleotides. A specific striking example of reentrant localization exhibiting the behavior in the parameter space has been confirmed by recent experiments \cite{Goblot2020}.
	
	\begin{figure}
		\centering 
		\includegraphics[width=0.50\textwidth, angle=-0]{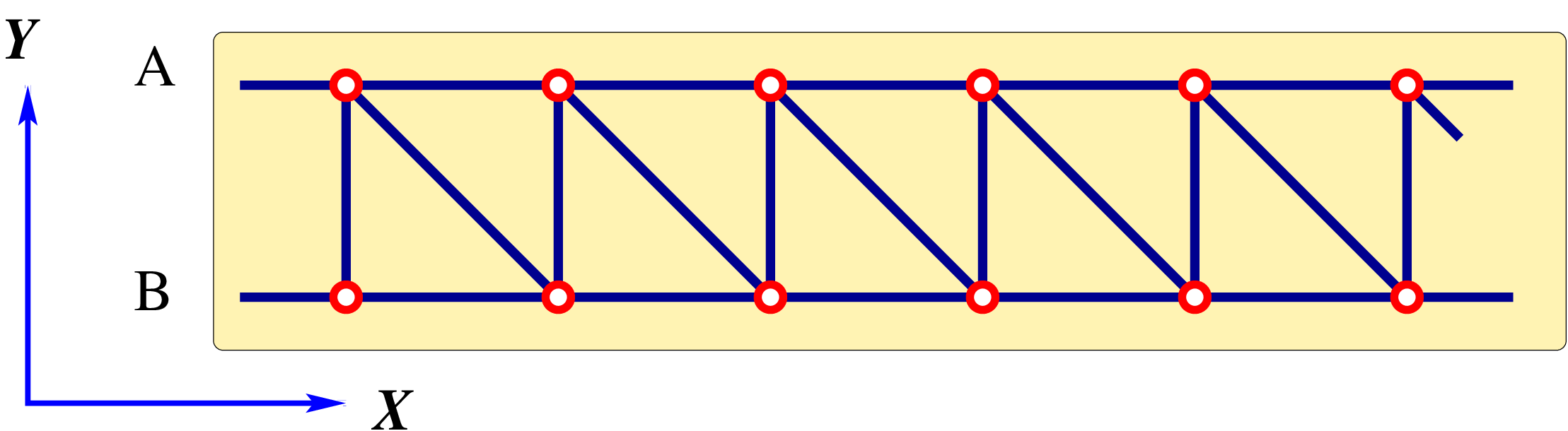}	
		\caption{A portion of a quasi-one dimensional zig-zag ladder.} 
		\label{lattice}%
	\end{figure}
	
	\section{Model System and Hamiltonian}
	We investigate the energy spectrum of a quasiperiodic two-port zig-zag ladder network as cited
	in the Fig.~\ref{lattice}. We embrace a tight-binding prescription, include nearest and next-nearest-
	neighbor hopping integrals $t$ and $\lambda_{NNN}$ inside a plaquette of the underlying
	ladder. The system is demonstrated by the Hamiltonian, viz.,
	
	\begin{equation}
		H_s
		=
		\epsilon_0 \sum_j d_j^\dagger d_j
		+
		\sum_{\langle j,k \rangle}
		t_{jk}
		\left[
		d_k^\dagger d_j + h.c.
		\right]
	\end{equation}
	
 Here, $d_j (d_j^{\dagger})$ are the annihilation (creation) operators respectively. A quadiperiodic Aubry-Andr{e} Harper
	(AAH) modulation is incorporated in a deterministic fashion in long-range overlap integral as,
	$\lambda_{NNN}=\lambda_0[1+\cos (\pi \beta n^{\alpha} a)]$, where $\lambda_0$ is the
	strength of the NNN connection and $\beta$ denotes the frequency of the modulation and $0 \le
	\alpha \le 1$ defines a slowness exponent controlling the variation. The intra-arm and inter-arm
	connections are taken as identical. With $\alpha=1$, fractional or irrational values of $\beta$
	essentially represents staggered or quasiperiodic variation of the long-range connection. For
	$\beta=(1+\sqrt{5})/2$, a Diophantine number, and $t=1$, we show that the aperiodically
	distorted zig-zag ladder hosts a reentrant localization-delocalization transition associated with a
	single-particle mobility edge (SPME) in the energy landscape. The interplay between the nearest
	neighbor hopping and long-range hopping may essentially invite an interesting spectral
	competition leading to a drastic modification of the spectrum. To characterize the nature of single
	particle eigenstates, we rely on the evaluation of inverse participation ratio (IPR) and the
	normalized participation ratio (NPR) which are often used as significant diagnostic tools. For the
	$n$-th normalized eigenstate $\psi_n$, IPR and NPR are defined as,
	
\begin{align}
\text{IPR}_n
&=
\sum_{j=1}^{N}
|\psi_n^j|^4,
\qquad
\text{NPR}_n
=
\left(
N
\sum_{j=1}^{N}
|\psi_n^j|^4
\right)^{-1}.
\end{align}
	
	In the thermodynamic limit, vanishing $IPR_n \propto (1/N)$ describes a signature of extended mode
	and for insulating phase, it reaches to unity. $NPR_n$, by definition, shows a contrasting
	behavior. Additional convenient strategy to track how the eigenstates evolve with the variation of different parameters, we also workout the averaged IPR and NPR over a subset of eigenstates,
	viz.,
\begin{equation}
\begin{aligned}
\langle \mathrm{IPR} \rangle
&=
\frac{1}{N}
\sum_{n=1}^{N}
\mathrm{IPR}_n,
\qquad
\langle \mathrm{NPR} \rangle
=
\frac{1}{N}
\sum_{n=1}^{N}
\mathrm{NPR}_n.
\end{aligned}
\end{equation}
	$\langle \text{IPR} \rangle \rightarrow 1$
	and
	$\langle \text{NPR} \rangle \rightarrow 0$
	for a completely localized state,
	in the asymptotic limit and vice-versa for extended one.
	For the critical phase, both the quantities take intermediate
	values. With the above knowledge, one may also workout
	the fractal dimension $D_n$ to probe the multifractality of
	the eigenspectrum, viz.,
	
	\begin{equation}
		D_n
		=
		-
		\lim_{N \to \infty}
		\frac{
			\log(\text{IPR}_n)
		}{
			\log(N)
		}
	\end{equation}
	
	This fractal dimension $D_n$ approaches towards unity for extended
	mode, zero for insulating mode and $0 < D_n < 1$ for critical mode.

	\section{Results and analysis}
	
\begin{figure}
		\centering 
		\includegraphics[width=0.45\textwidth, ]{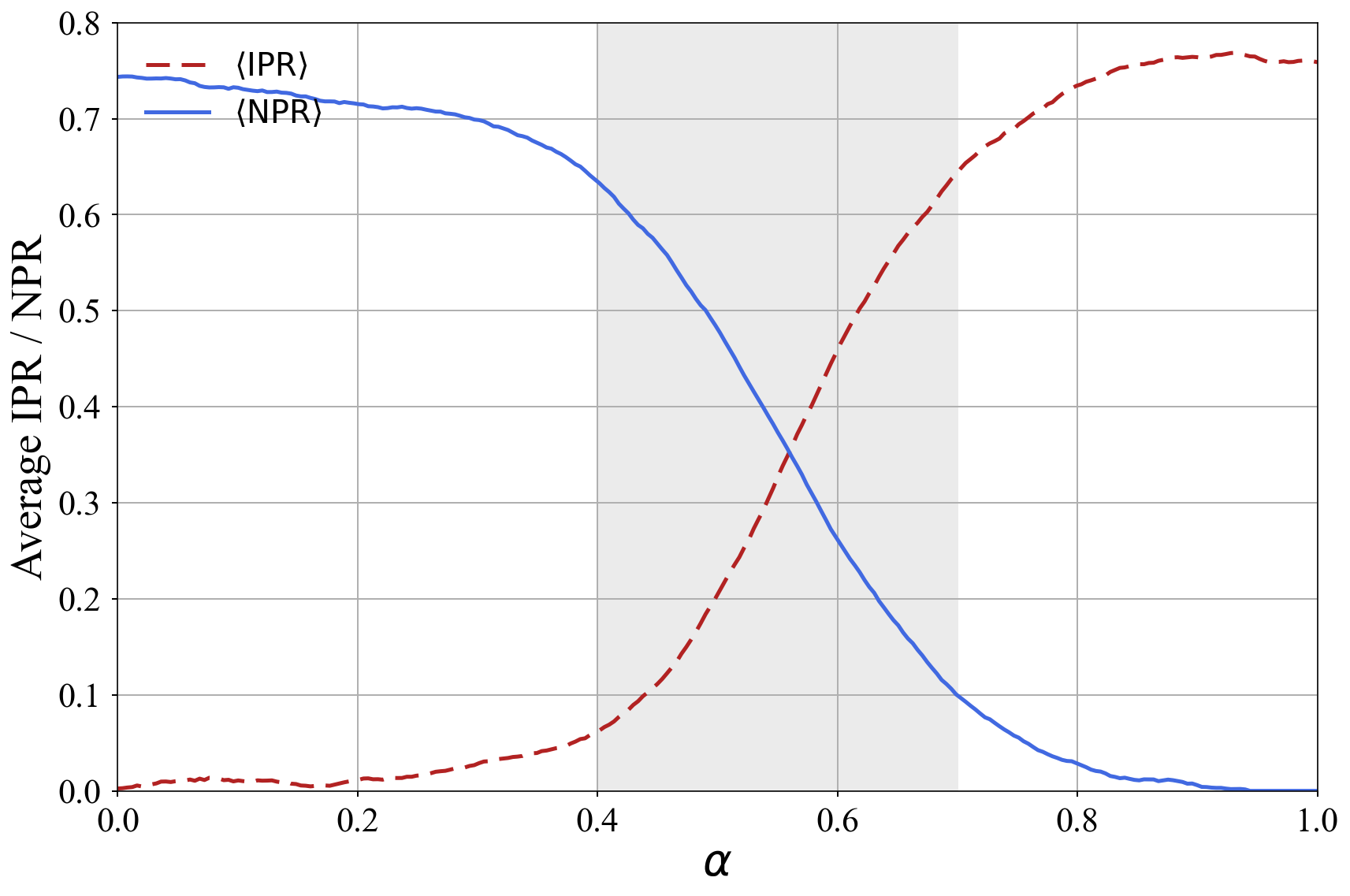}	
		\includegraphics[width=0.45\textwidth, ]{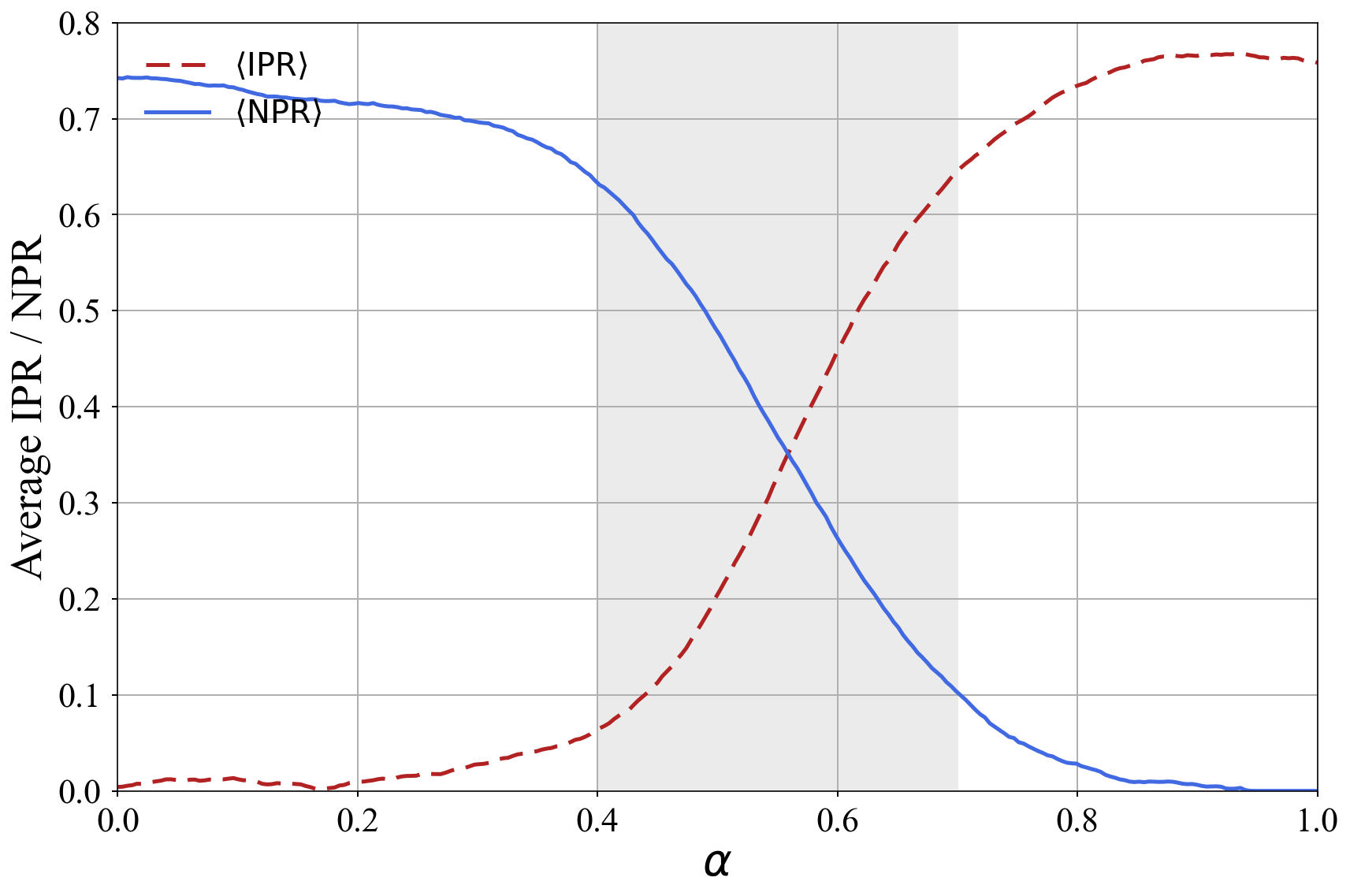}	
		\caption{(a) and (b) The average inverse participation ratio, $\langle \mathrm{IPR} \rangle$, and the average normalized participation ratio, $\langle \mathrm{NPR} \rangle$, with respect to $\alpha$ respectively, for disorder strengths $\lambda_0<t$ \& $\lambda_0>t$, where $t$ is the vertical hopping.}
		\label{avgiprnpr}%
	\end{figure}

	It is well-known that pure AA model exhibits a localization transition without any SPME. In this
	analysis, we demonstrate that the localization transition happens through a critical regime
	hosting the SPME for both $t  < \lambda_0$ and $t > \lambda_0$.
    To unfold this interesting peculiarity, we have cited the variation of $\langle$IPR$\rangle$ and $\langle$NPR$\rangle$ with the slowness factor $\alpha$ in the Fig.~\ref{avgiprnpr}. Here the average of IPR and NPR are calculated by considering all the eigenstates for definite value of $\alpha$. It is appreciating to report that the variations show a mixed regime where both $\langle$IPR$\rangle$ and $\langle$NPR$\rangle$ are finite for both $t \langle \lambda_0$ and $t \rangle \lambda_0$. This implies that both extended and localized modes coexist for that particular range of $\alpha$ (shaded portion). It is seen that this shaded portions of the plots showing the crossover of average IPR and NPR are no longer sensitive on the relative strength of the nearest neighbor hopping and next nearest neighbor hopping. Mixed phase is entirely the consequence of the slowness index $\alpha$. This region confirms the existence of critical phase in the spectrum exhibiting the SPME. The localization transition and the associated SPME induced by the modulation of the slowness factor $\alpha$ are subsequently discussed with the IPR intensity plot, phase diagram and multifractal analysis. It is well-known that in quasiperiodic networks exhibiting localization transition hosting the SPME, once the system enters in the non-conducting phase, it monotonically remains in that phase with respect to the strength of the aperiodicity. Unlike to this description, the slowness index here offers an intermediate regime where criticality appears in the spectrum and this happens irrespective of the strength of the quasiperiodic long-range overlap integral.
		
	\subsection{Re-entrant localization-delocalization}
	
	\begin{figure*}[ht]
		\centering
		(a)\includegraphics[width=0.45\textwidth]{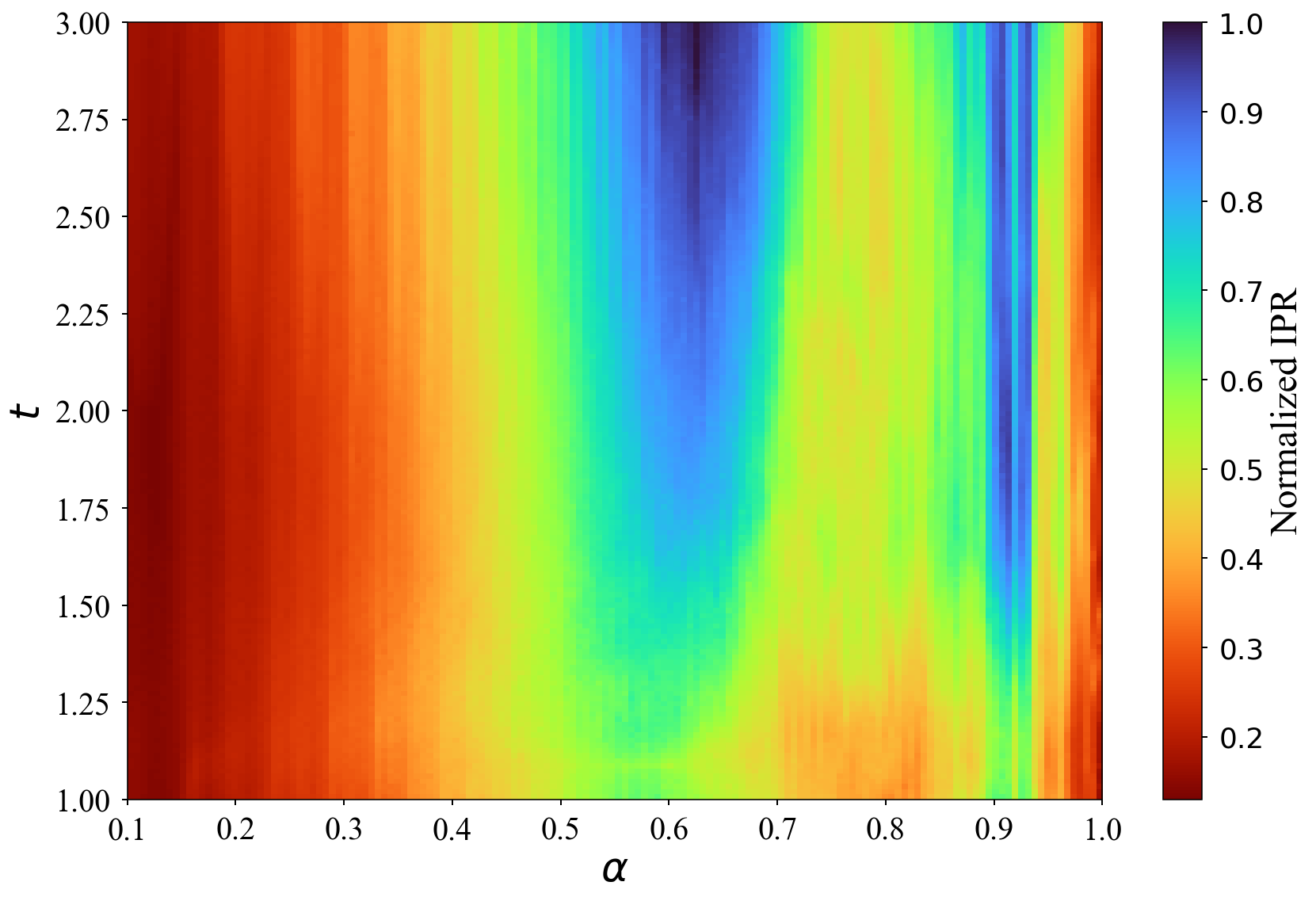}
		(b)\includegraphics[width=0.45\textwidth]{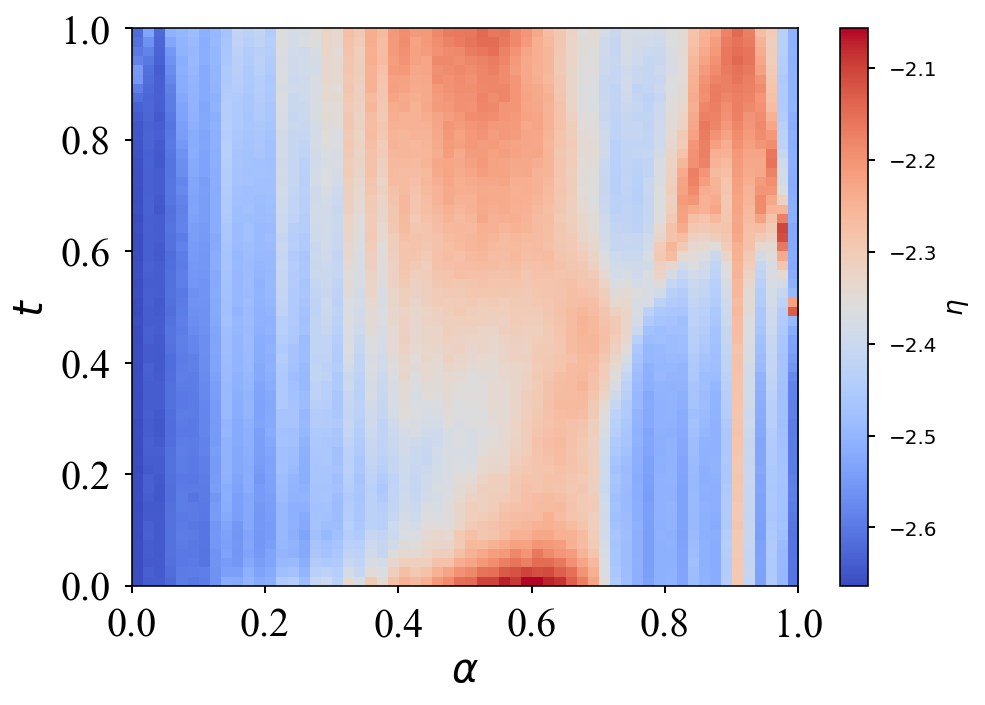}
		(c)\includegraphics[width=0.45\textwidth]{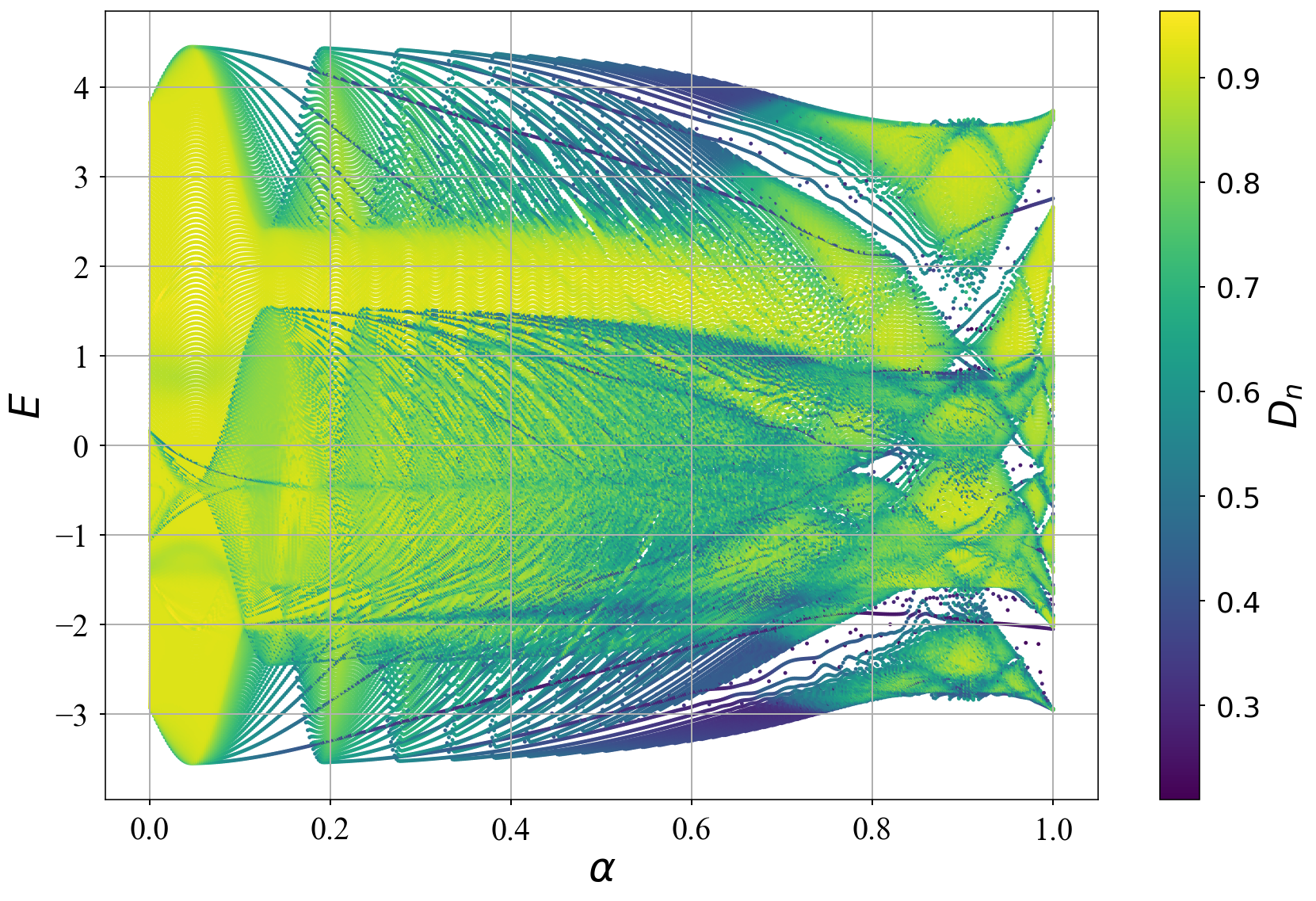}
		(d)\includegraphics[width=0.45\textwidth]{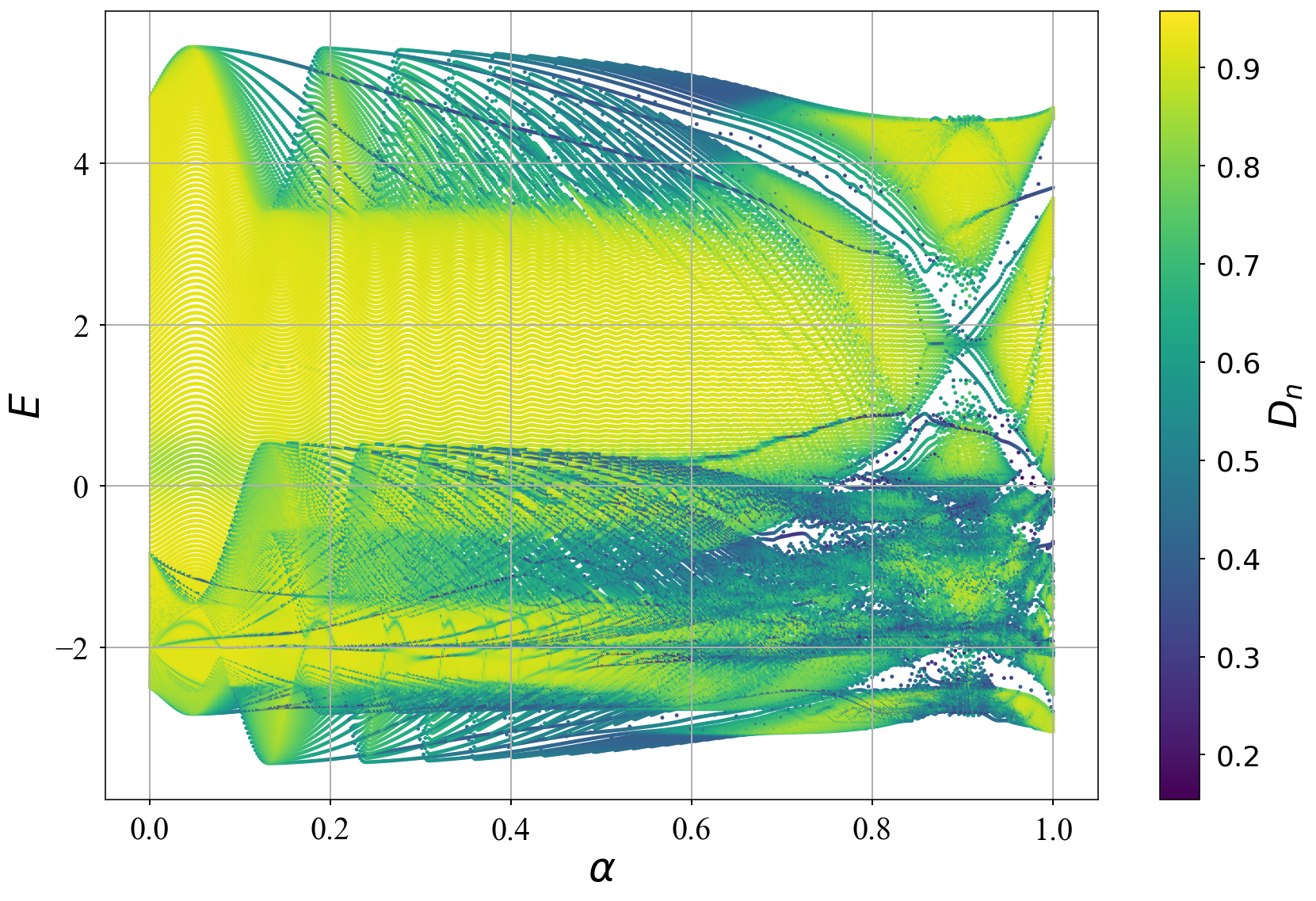}

		\caption{(a) IPR intensity distribution in $t$ vs $\alpha$ space, (b) phase diagram of $\eta$ in $t$ vs $\alpha$ space, and fractal dimension in $E$ vs $\alpha$ plane with (c) $t \langle \lambda_0$ and (d) $t \rangle \lambda_0$.}
		\label{renloc}
		
	\end{figure*}
	The IPR intensity plot in the $t-\alpha$ space
	(Fig.~\ref{renloc}) gives a clear indication regarding the re-entrant localization-delocalization
	transition. The deep red patch populated by the states having vanishingly small IPR at the left
	signifies the extended states. The system then encounters a critical regime for moderate slowness
	factor. As we go beyond this, we notice the extended states reappear in the spectrum. It is to be
	noted that the re-entrant transition is not visible for a system supporting monotonic disorder (i.e., random scatterers). Instead, occurrence of re-entrant localization-delocalization transition is here
	entirely controlled by the interplay of the kinetic parameters of the Hamiltonian. Another
	standard manifestation of the key results is the representation of re-emerging transition through
	the phase diagram in the $t-\alpha$ plane. The phase plot is obtained by evaluating a quantity
	$\eta$ \cite{Li2020} as,
	\begin{equation}
		\eta=\log_{10}[\langle IPR \rangle \times \langle NPR \rangle]
	\end{equation}
	The presence of the insulating regime (marked by red colors) becomes prominent because of the
	intermediate value of $\alpha$. The delocalized phase (marked by blue color) again reappears for moderate $\alpha$ values as we go beyond $\alpha=0.6$ after the first blue patch. We
	complement these findings by stating that the slowness index of the off-diagonal modulation in
	its aperiodic limit thus controls the localization of the wavepacket in this zig-zag ladder
	geometry and hence the non-trivial re-entrant transition associated with the SPME are ascribed to
	the said spectral competitiveness. The density plot of fractal dimension $D_n$ in the $E-\alpha$
	plane (Fig.~\ref{renloc}) also reflects the localization transition and shows that a subset of the allowed
	eigenstates changes the phase from conducting to insulating readily allowing a SPME. The
	extended states appear for extremely low values of the slowness index $\alpha$.
	As $\alpha$ increases, the landscape enters into a critical region where the aperiodic distortion
	induced by the AAH off-diagonal modulation begins to compete with the kinetic delocalization.
	This spectral interplay leads to a non-trivial coexistence of diffusive and non-diffusive modes
	within the same spectrum, consequently giving rise to a SPME. For some moderate to larger
	value of $\alpha$, delocalized states reappear in the energy spectrum.

	
	\subsection{Quantum dynamical perspective}
	
	\begin{figure}[t]
		\centering
		\includegraphics[width=0.45\textwidth]{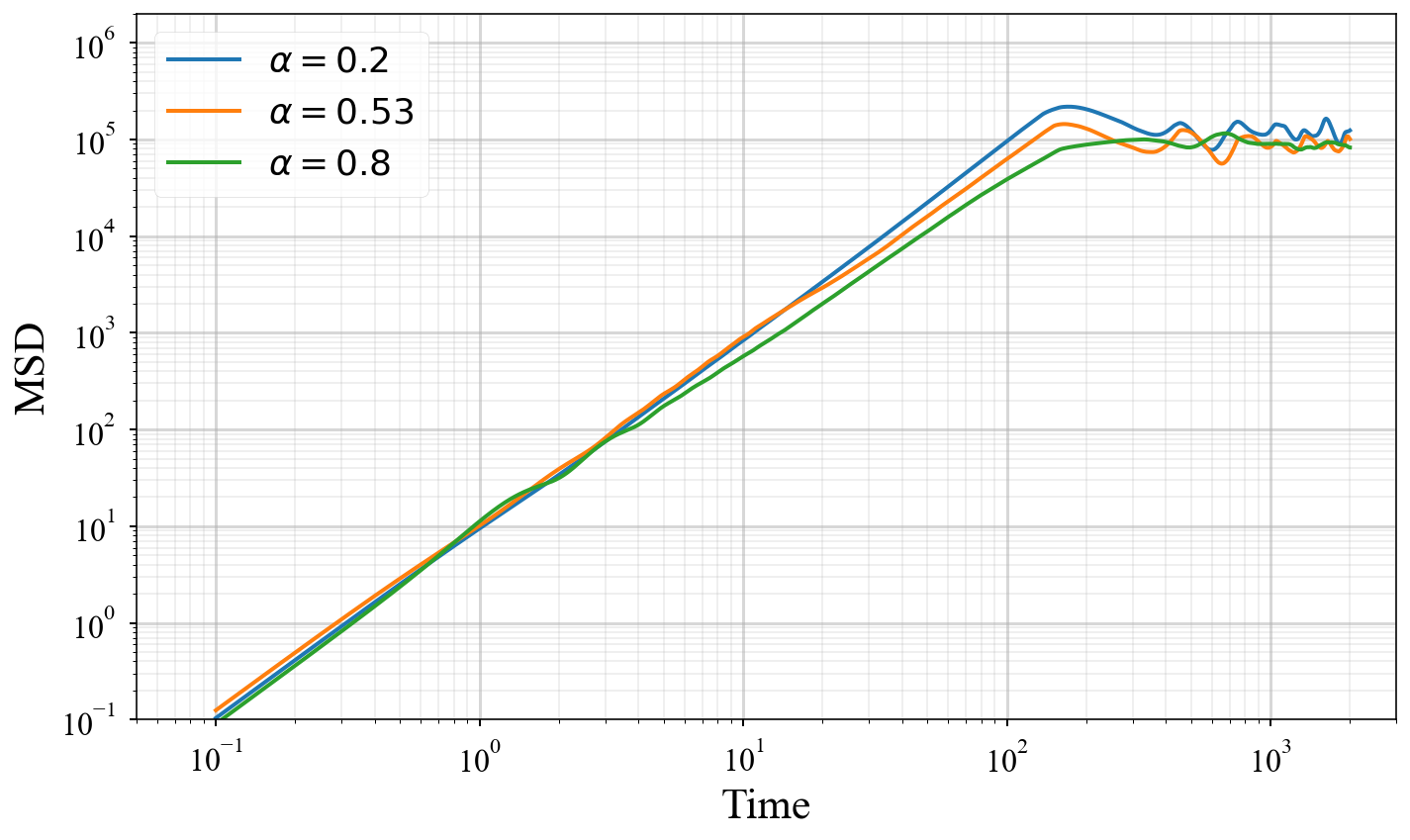}
		\caption{Time evolution of the mean square displacement (MSD) for three representative values of the slowness exponent $\alpha$ with the quasiperiodic modulation parameter fixed at the golden mean, $Q=(1+\sqrt{5})/2$. The different transport regimes are characterized by the corresponding transport exponent $\mu$.}
		\label{msd}
	\end{figure}
	
	In this section, we investigate the dynamical properties~\cite{Katsanos1995} of the zig-zag ladder with quasiperiodic (AAH) long-range hopping. The real-time evolution of an initially localized wave packet is studied for the quasiperiodic modulation parameter fixed at the golden mean, $\beta =\frac{1+\sqrt{5}}{2}$ and for three representative values of the slowness exponent $\alpha$. The choice of these values is motivated by the re-entrant localization transition observed in the phase diagrams in Fig~\ref{renloc}. The dynamical behavior is characterized through the time evolution of the mean square displacement (MSD), from which we extract the transport exponent.
	
	At time $t=0$, the wave packet is expressed as
	\begin{equation}
		|\Psi(0)\rangle
		=
		\sum_n C_n(0)|\psi_n\rangle,
	\end{equation}
	where $|\psi_n\rangle$ denotes the eigenstate of the Hamiltonian and $C_n(0)$ is the initial probability amplitude. When the particle is initially localized at the $m$th lattice site,
	\begin{equation}
		C_n(0)
		=
		\langle \psi_n|m\rangle .
	\end{equation}
	
	The time-evolved wave function is, therefore, given by
	\begin{equation}
		|\Psi(t)\rangle
		=
		\sum_n
		C_n(0)e^{-iE_nt}
		|\psi_n\rangle,
	\end{equation}
	where $E_n$ represents the corresponding eigenenergy.
	
	The transport properties are quantified by the mean square displacement (MSD),
	\begin{equation}
		\mathrm{MSD}(t)
		=
		\sum_i
		(i-m)^2
		|\psi_i(t)|^2,
	\end{equation}
	which follows the power-law scaling relation
	\begin{equation}
		\mathrm{MSD}(t)\sim t^{\mu},
		\label{msd_scaling}
	\end{equation}
	where $\mu$ is the transport exponent. It is obtained from the slope of the $\log(\mathrm{MSD})$ versus $\log(t)$ plot.
    The calculated transport exponents for the three representative values of the slowness exponent are
	\[
	\mu = 2.057615,\qquad
	\mu = 0.024963,\qquad
	\mu = 1.884032,
	\]
	From the value of the exponent $\mu$ we can charecterize if the system is localised or diffusive or super diffuse or ballistic. corresponding to $\alpha=0.2$, $\alpha=0.53$, and $\alpha=0.8$, respectively. The first and third cases exhibit ballistic and super diffusive transport which can also be confirmed from the phase transition diagrams. In contrast, the nearly vanishing transport exponent at $\alpha=0.53$ indicates an almost complete suppression of transport, consistent with a localized dynamical regime where the wave packet remains confined around its initial position. This behavior provides independent dynamical evidence for the re-entrant localization transition inferred from the IPR / NPR or t vs alpha plots. 
	
	Overall, the dynamical signatures observed for the quasiperiodic system are in excellent agreement with the phase transition characteristic discussed earlier. The coexistence of ballistic-like spreading and  suppressed transport indicates the re-entrant localization transition induced by the quasiperiodic long-range hopping in the zig-zag ladder.

	\section{Single-particle mobility edge}
    \subsection{Multifractal analysis}
	
	Multifractal behavior is a basic characteristic of strongly fluctuating wave functions at criticality. Simple monofractal analysis considers a single scaling exponent across all scales, but multifractal analysis uses a set of exponents presenting the scaling of moments of some probability distribution. For the case of electron localization, the standard measure is the moment of wavefunction amplitude $|\psi(r)|^{2q}$. We therefore define generalized inverse participation ratio ($I_q$) as the moments of eigenstate intensities,
	\begin{equation}
		I_q = \int |\psi(r)|^{2q} d^d r
	\end{equation}
	$I_q$ maintains a standard power-law fashion with the system size $N$ as, $I_q\propto N^{-\tau_q}$. The exponent $\tau_q$ is thus defined as,
	\begin{equation}
		\tau_q = -Lt_{N \rightarrow \infty} \left(\frac{ln I_q}{ln N}\right)
	\end{equation}
	The generalized fractal dimension $D_q$ defined through $\tau_q = D_q (q-1)$ quantifies the multifractal behavior. 
	It is to be noted that the critical exponent $\tau_q$ (a) follows linearity with $D_q= d$ for metals, (b) becomes flat in localized systems with $D_q= 0$, and (c) shows a nonlinear fashion as a function of $q$  at critical points.

	As we observe that the spectral landscape offers two distinct regimes of energy with different quantum phases for aperiodic long range hopping integral. To unravel the multifractality of the wavefunctions we use the standard multifractal analysis. It is to 
	be noted that
	this is a valid way-out which has been successfully applied in certain
	disordered and quasiperiodic tight-binding models. 
	As presented in the Fig.~\ref{renloc}, the states residing in the extended regimes have extremely low
	value (~$10^{-3}$) of IPR while for the insulating zone, the states occupy relatively larger value
	(~$10^{-1}$) of IPR. This relative change of IPR values eventually indicates that there exist
	mobility edges in the eigenspectrum. A finite-size scaling analysis with the help of ln IPR vs. ln
	N plot shown in the Fig.~\ref{renloc} also confirms our claim. As it is well-known that $IPR\propto N^{-
		\delta}$, N being the dimension of the Hamiltonian matrix. Here, $\delta=1$ denotes the
	extended state and small but finite value of $\delta \ (0 < \delta < 1)$ describes the insulating
	phase. For any energy belonging to the extended zone, IPR goes as 1/N indicating nonzero
	spread of the wavefunction envelope over the entire lattice. The spatial spread is confined over
	few sites for the insulating eigenenergy. We have picked up selective energies from the two
	distinct regimes and plotted it. The dependence of ln IPR with the system size distinctively cites
	the coexistence of two different quantum phases.
	
	To perform the multifractal analysis, we set $\beta = (\sqrt{5}+1)/2$ and $\alpha = 1/2$ and carefully pick up two energies eigenvalues $E= 1.75$ and $5.75$ 
	from the extended and non-conducting regimes based on their IPR values. The variation of the
	generalized inverse participation ratio (GIPR) as a function of the system size describes the
	difference between the two phases. It is seen that $I_q$ changes nonlinearly with $N$ with the
	effect becoming prominent at higher values of $q$ while, on the contrary, the variation is linear
	at all $q$ for extended phase. We also calculate the multifractal exponent $\tau_q$ for these two
	states. The linear fit (red curve) indicates the resonant character of the state and for the insulating
	phase exponent follows a non-linear fashion. This non-linear trend becomes sharper for higher
	value of $q$. This is a clear signal of multifractality of the spectrum.

\begin{figure*}[t]
\centering

(a) \includegraphics[width=0.3\textwidth]{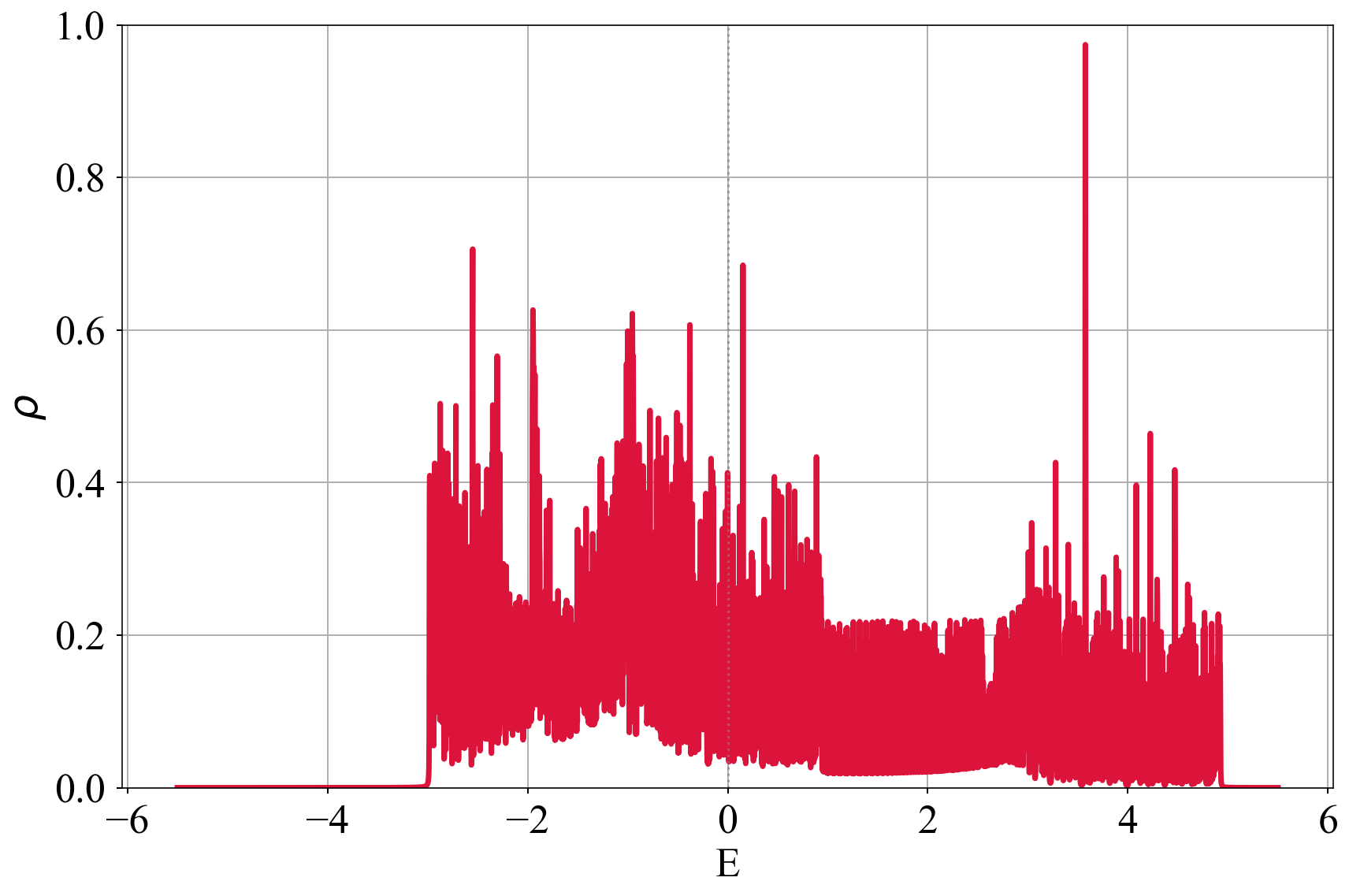}\hfill
(b) \includegraphics[width=0.3\textwidth]{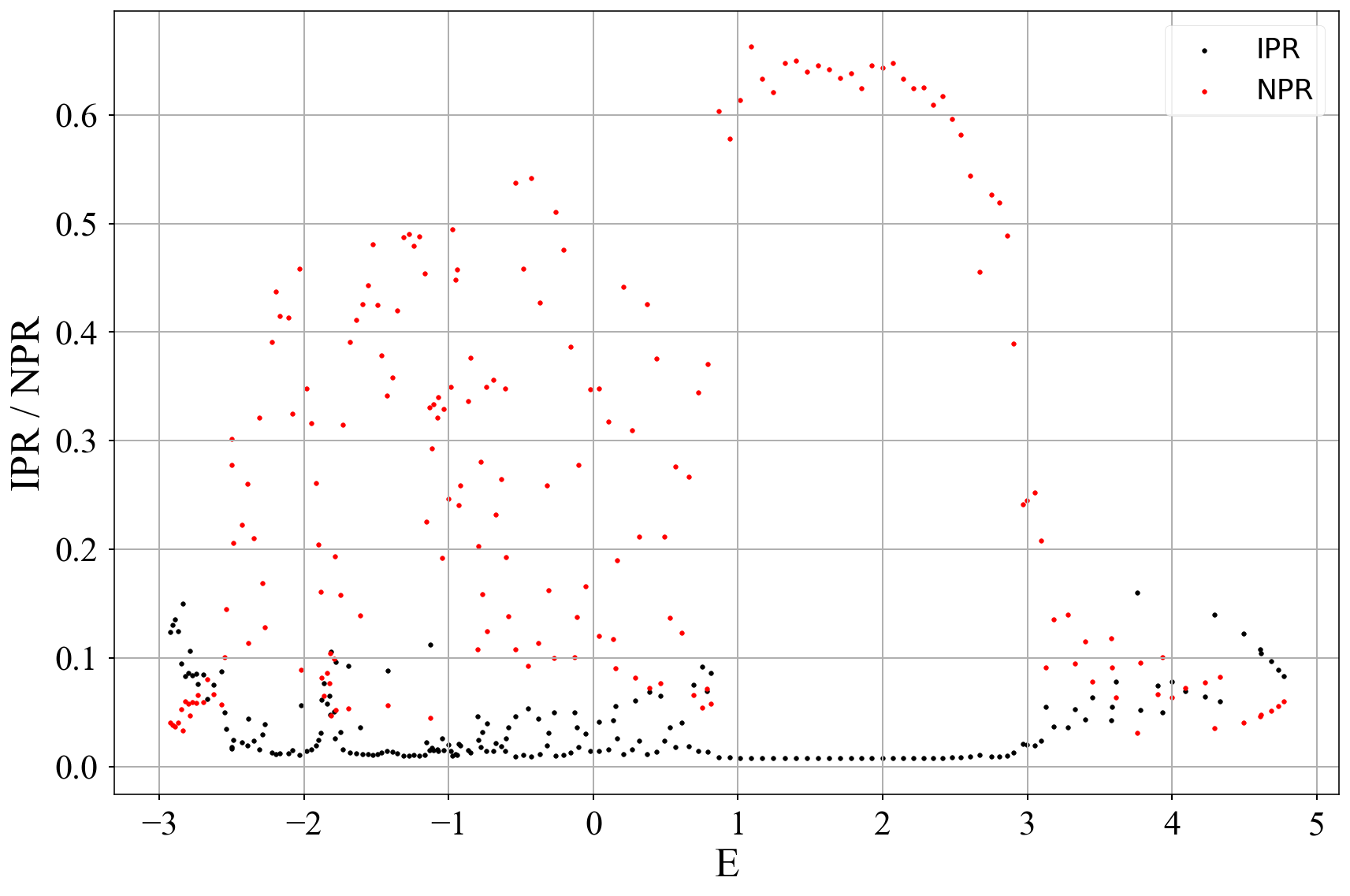}\hfill
(c) \includegraphics[width=0.3\textwidth]{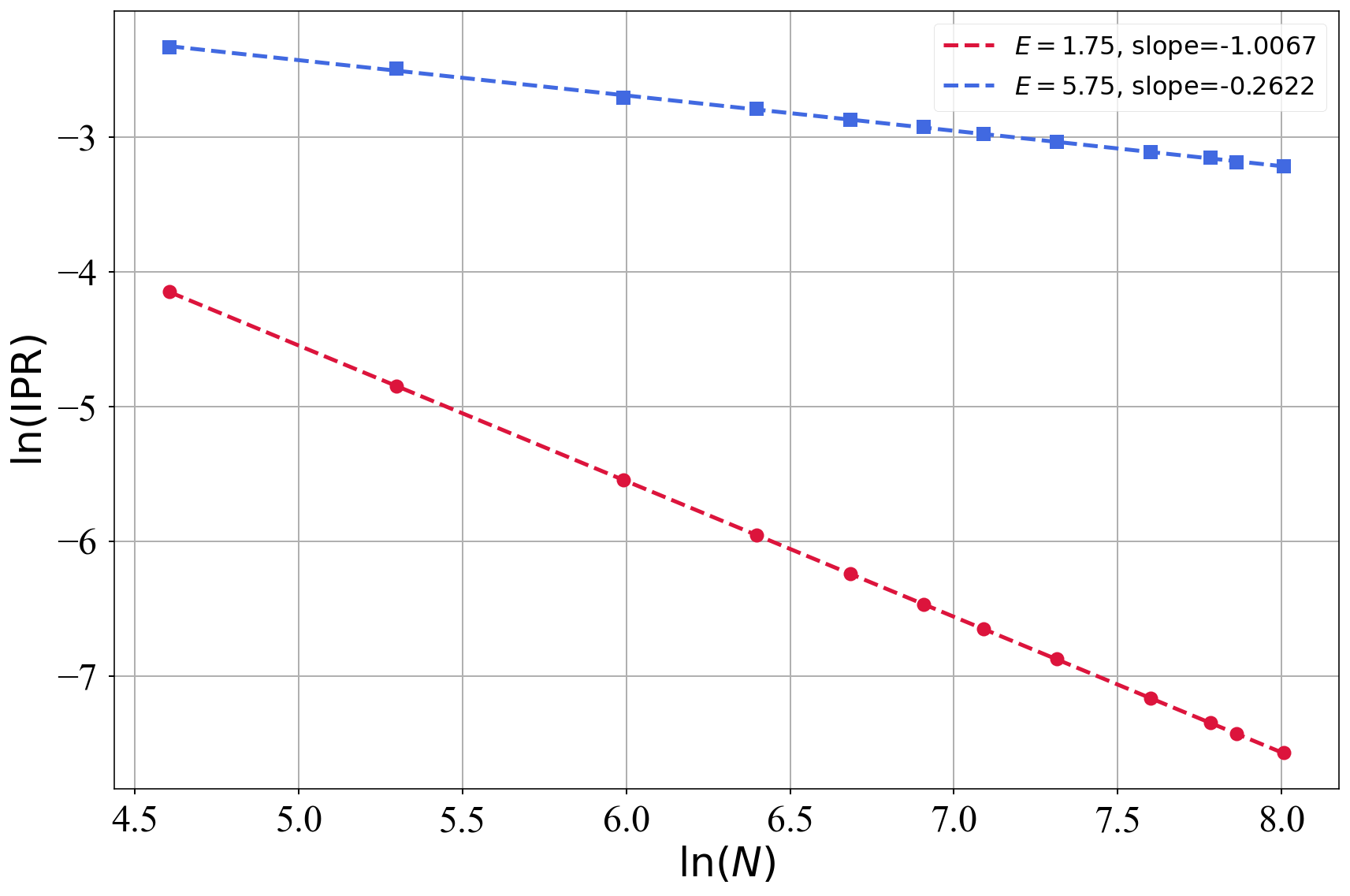}

\vspace{0.4cm}

(d) \includegraphics[width=0.3\textwidth]{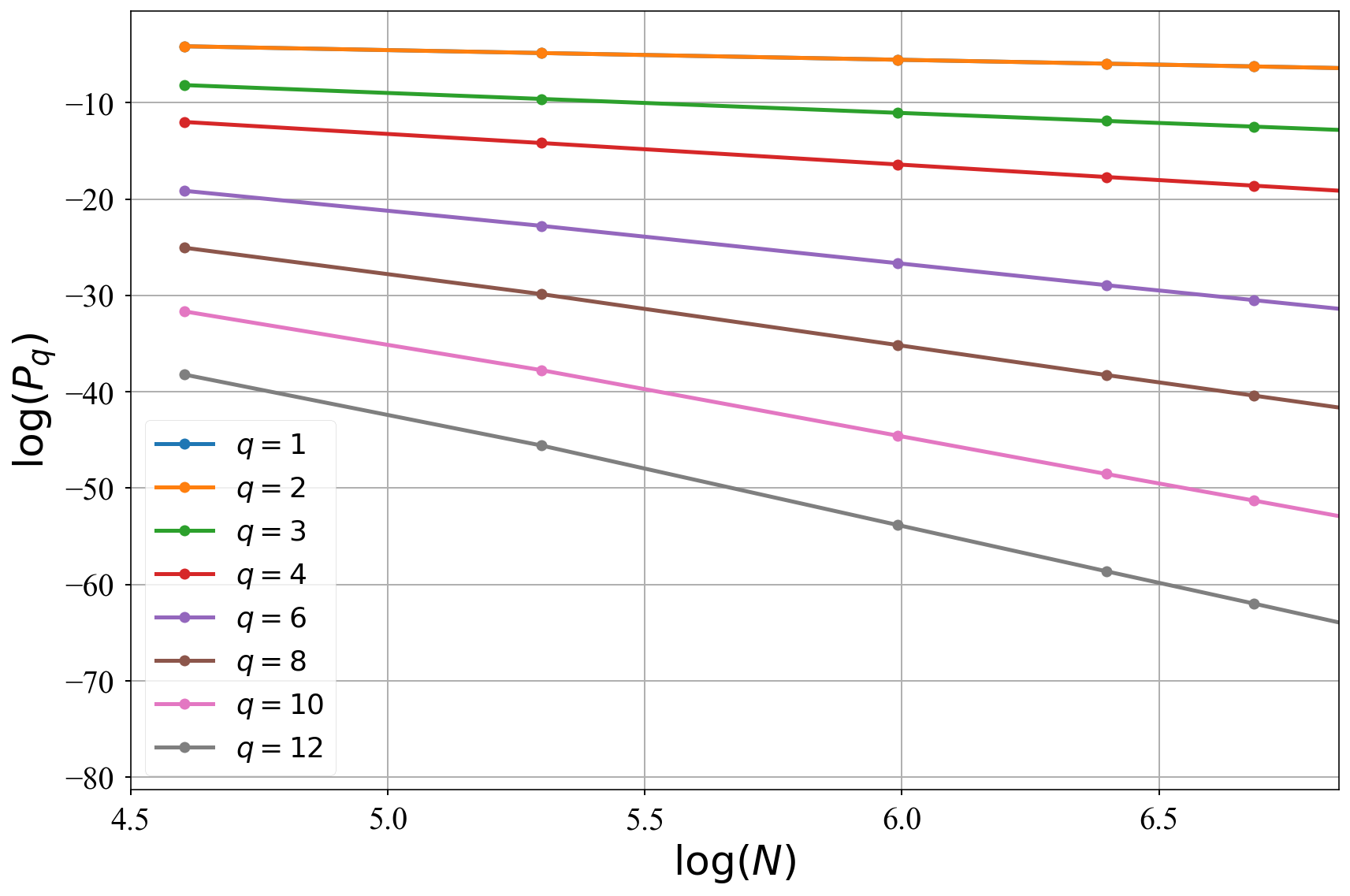}\hfill
(e) \includegraphics[width=0.3\textwidth]{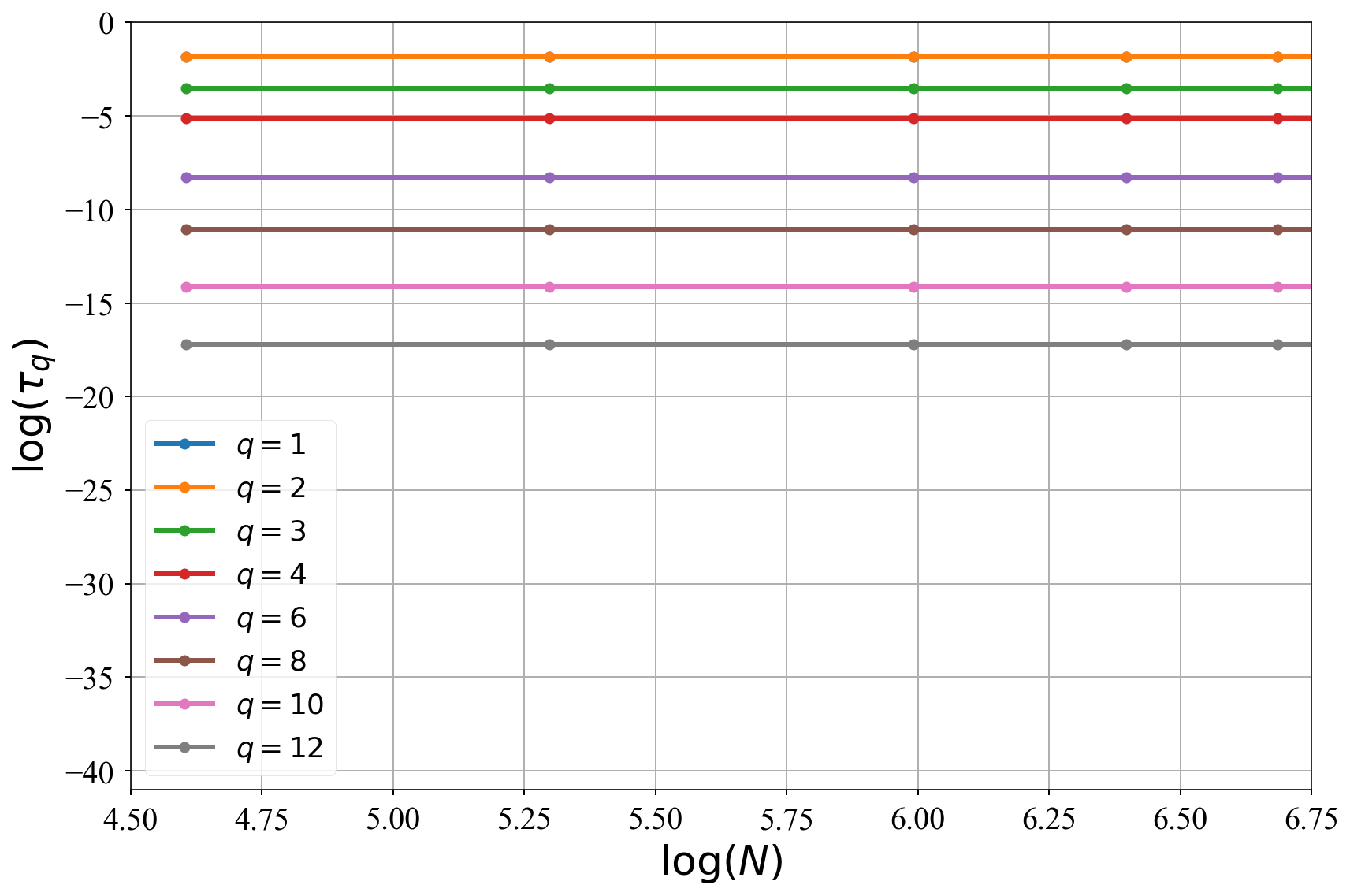}\hfill
(f) \includegraphics[width=0.3\textwidth]{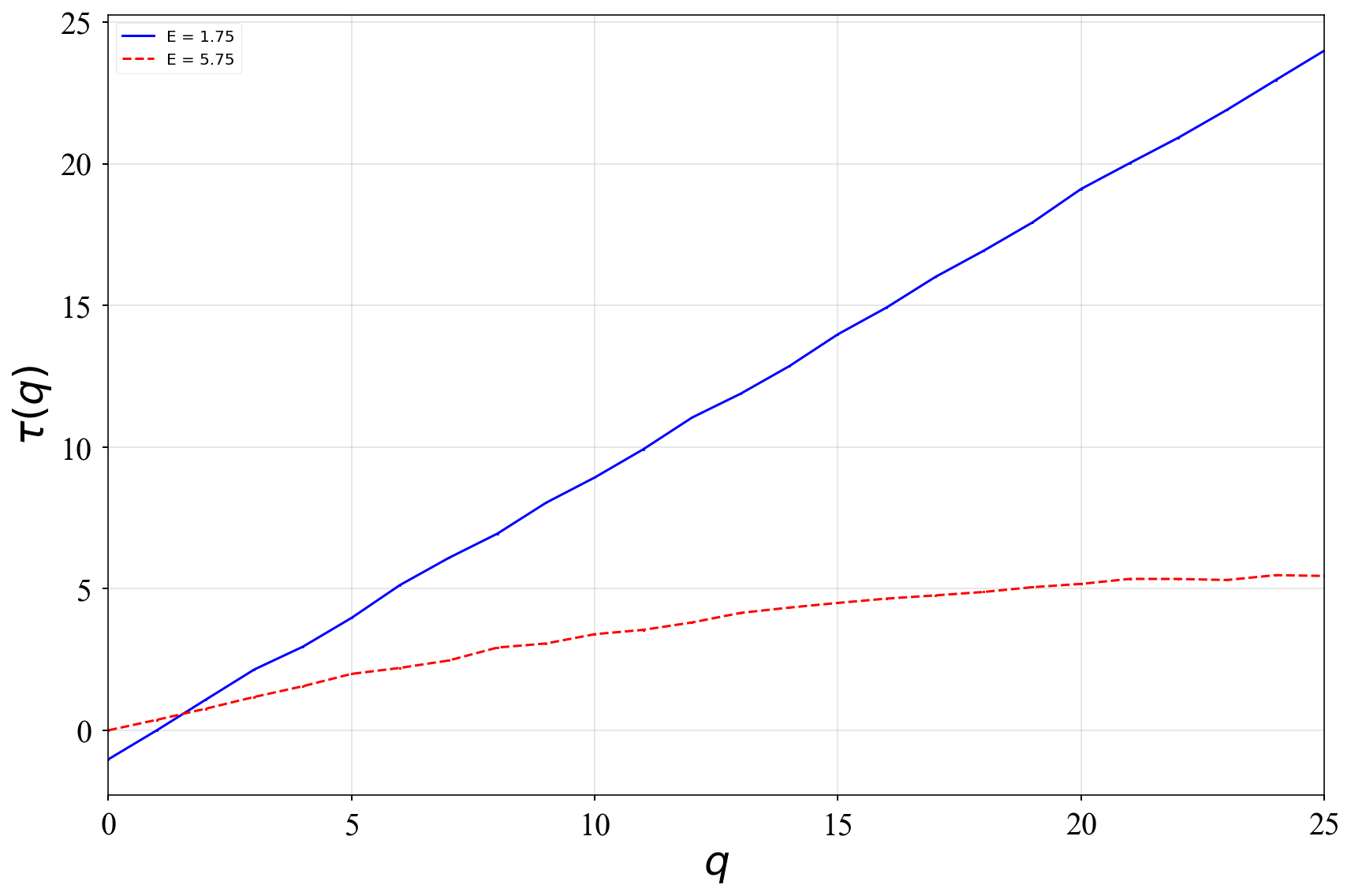}

\caption{(a) Variation of density of states with energy (b) inverse participation ratio with energy (c) variation of IPR with system size,
(d)-(f) MFA spectrum for $E=1.75$ (extended state) and $E=3.75$ (critical state). The finite-size scaling yields slopes of $-1.0067$ for the extended state and $0.2622$ for the critical state.}
\label{mfa}
\end{figure*}
	\subsection{Quantum Butterfly} 

    	\begin{figure}[ht]
		\centering
		
		\includegraphics[width=0.45\textwidth]{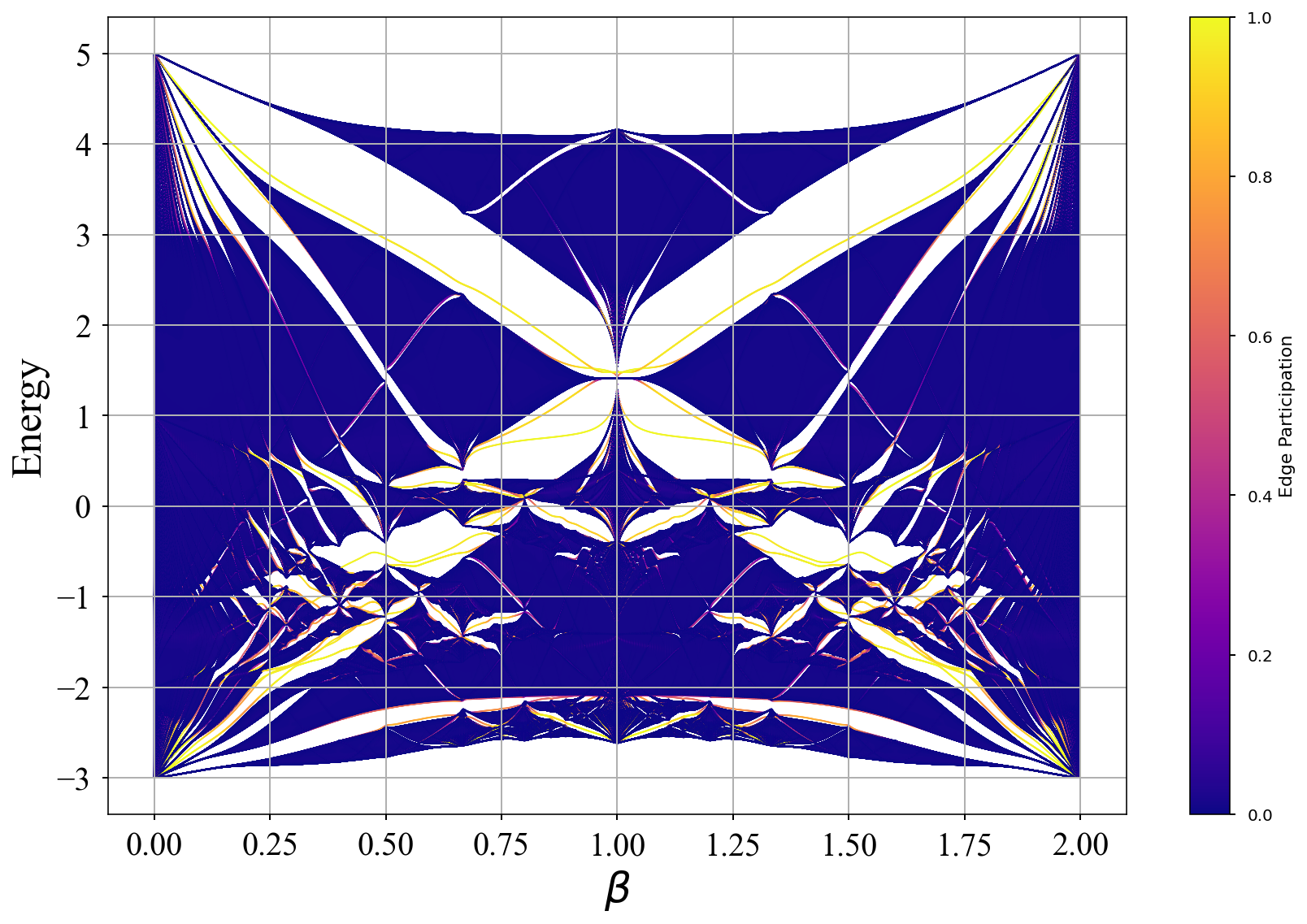}

		\caption{Quantum butterfly spectrum for zig-zag ladder with aperiodic kinetic modulation.}

		\label{butter}
		
	\end{figure}
\begin{figure}[ht]
    \centering

    (a)\hspace{0.5em}
    \includegraphics[width=0.38\textwidth]{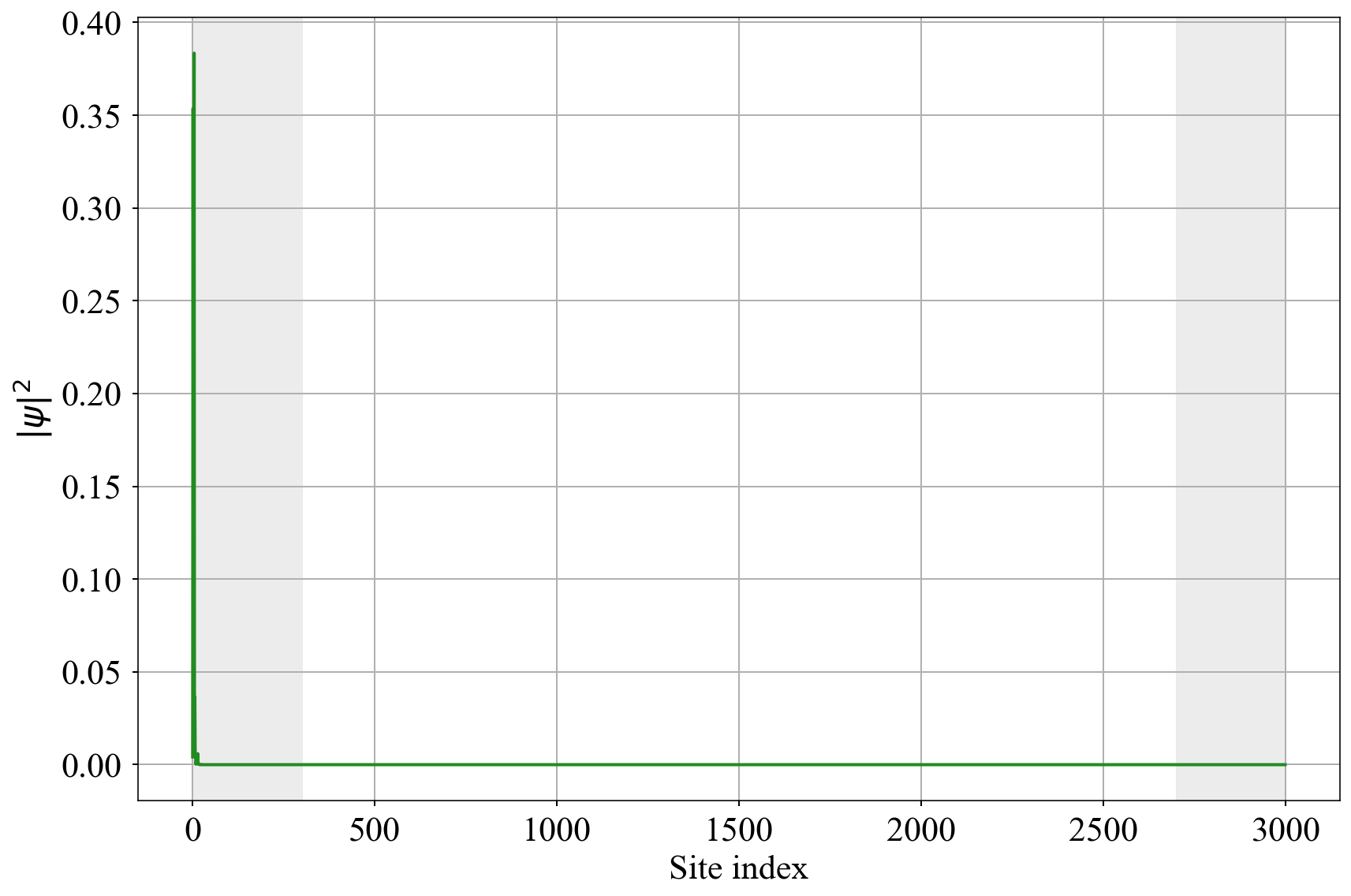}
    \hfill
    (b)\hspace{0.5em}
    \includegraphics[width=0.38\textwidth]{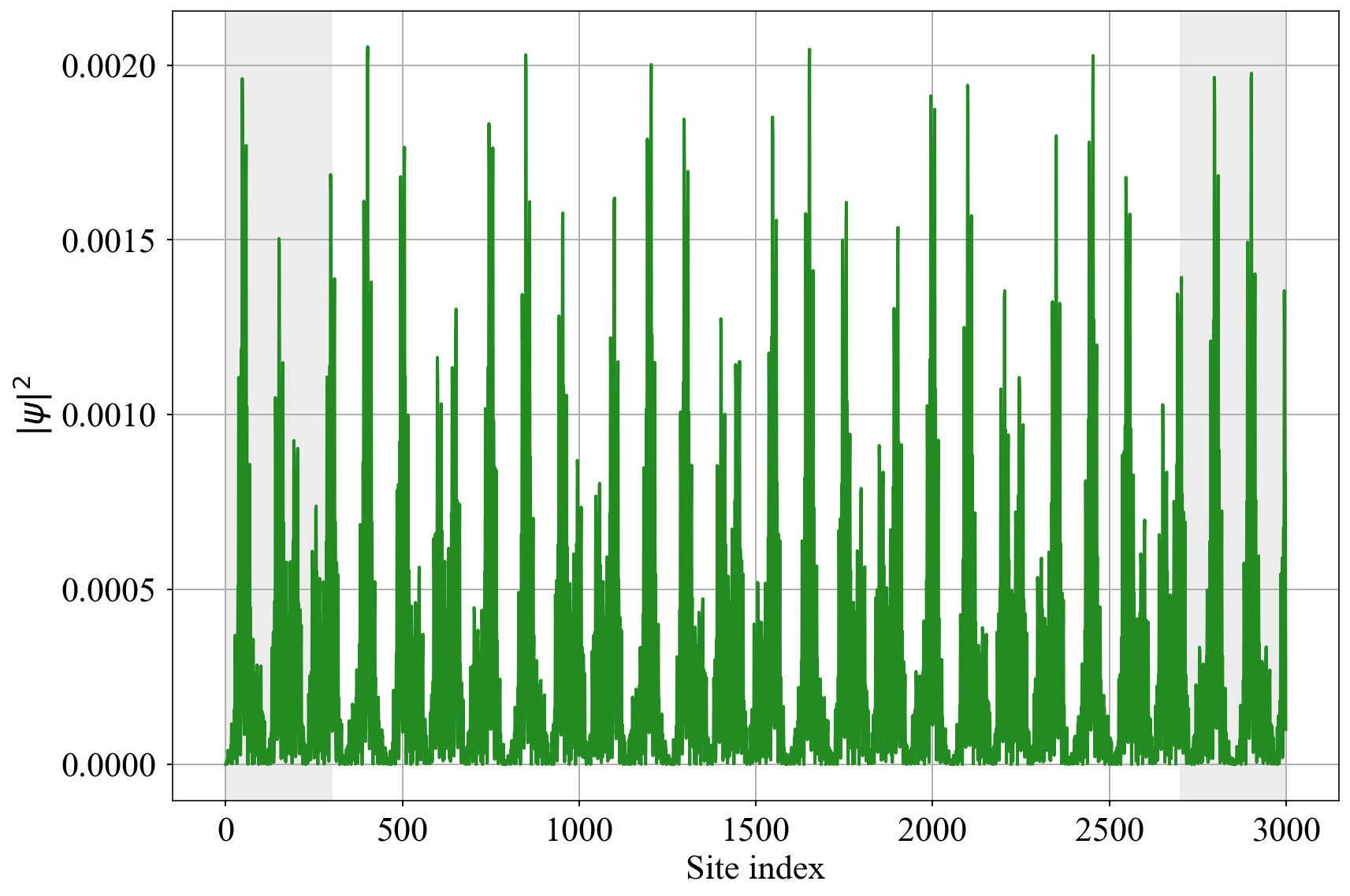}

    \caption{Probability density distributions corresponding to (a) $E=4.75$ and (b) $E=2.75$.}
    \label{edge}
\end{figure}
	
	Before ending this discussion, we will demonstrate the overall multifractal eigenspectrum offered by the ladder. For this, we first study the energy spectrum as a function of the frequency $\beta$ of the off-diagonal modulation for different strengths $\lambda_0$. We set $t = 1.0$ without any loss of generality. As we notice that the variation with respect to $\beta$ produces quantum butterfly in the allowed eigenspectrum. The strength and the frequency of the modulation are responsible for creation of the self similar energy landscape. Exact diagonalization of the Hamiltonian reveals the eigenspectrum presented in the Fig~\ref{butter}. The dimension of the Hamiltonian matrix taken for this evaluation is $3000 \times 3000$.
	
	With minimal strength of aperiodic long-range 
	connectivity and $\alpha=1$,
	a quasi-continuous spectral
	profile starts to appear
	with minigaps separating the subbands. Band
	overlap near the band edge is also characterized in the spectrum.
	As we increase the strength further, it is seen that the gaps become more
	prominent and the spectrum takes a complete butterfly
	shape. (Fig.~\ref{butter}) The energy levels splits into a number
	of subbands for $\lambda_0=2$ and multifractal pattern appears
	which is an intriguing signature of the off-diagonal modulation
	imposed. This picture is very much identical to the
	Hofstadter butterfly \ref{} and hence can be considered as one
	of its interesting variant. This multifractal
	energy landscape disappears at the moment we fix a fractional
	slowness index
	$\left(
	\alpha=\frac{1}{2}
	\text{ in our case}
	\right)$,
	as demonstrated in the Fig~\ref{butter}.

    	\subsection{Edge modes}
	
	Edge states are electronic states whose wave-functions are predominantly confined near the boundaries of a finite size system. These staes possess a significant probability density at the edges which is different from the states we find in bulk states. The formation and robustness of these boundary states are dependent on the lattice geometry, hopping amplitudes and the nature of the external perturbation. In quasiperiodic systems, non-uniform hopping amplitudes can generate robust edge-localized modes through interference effects. To investigate the nature of those eigenstates, we examine the probability density $|\psi_i|^2$ of selected eigenstates corresponding to representative energies of the butterfly spectrum. The edge localization is quantified through the edge probability,
	
	\begin{equation}
		P_{\mathrm{edge}}=\sum_{i\in \mathrm{edge}} |\psi_i|^2,
	\end{equation}
	
	where the summation is performed over the boundary sites of the lattice. It can be characterized by  $P_{\mathrm{edge}}$ called edge weight. The states with large  are identified as edge-localized, whereas states with small edge weight correspond to bulk states. 
	
	To identify the boundary-localized states, we first compute the edge probability $P_{edge}$ for every eigenstate and map it onto the quantum butterfly, as shown in Fig.~\ref{butter}. The bright yellow areas shows the eigenstates  with large edge probability, whereas the dark branches represent bulk states. Representative energies are then selected from these two distinct regions to analyze their real-space probability distributions. The associated probability density profiles clearly distinguish the two types of eigenstates in Fig~\ref{edge}. Our numerical analysis reveals the existence of boundary-localized eigenstates whose probability density is strongly concentrated near one edge of the ladder while remaining negligible in the bulk. In contrast, the bulk states are distributed throughout the entire lattice. This demonstrates that the quasiperiodic AAH hopping modulation supports the formation of robust boundary modes. 
	
	To examine their stability, random disorder is introduced into the quasiperiodic AAH hopping amplitudes according to
	\begin{equation}
		\lambda_i \rightarrow \lambda_i+\delta_i,
	\end{equation}
	where $\delta_i$ is uniformly distributed in the interval $[-W_d/2,W_d/2]$. The edge modes remain localized up to a critical disorder strength of approximately $W_d^{c}\approx1.72$, beyond which the boundary localization is destroyed due to hybridization with the bulk states.
	
	These results indicate that the boundary-localized states are robust against moderate disorder in the quasiperiodic AAH hopping. The existence of a finite critical disorder strength suggests that the quasiperiodic hopping profile plays a crucial role in stabilizing the edge localization and provides an effective means to tune the boundary modes.

    \section{Closing remarks}
    In conclusion, we have unraveled the interesting spectral features of a quasi-one dimensional zig-zag ladder with quasiperiodic next nearest neighbor hopping integral within a tight-binding framework. The interplay between the kinetic parameters of the Hamiltonian essentially determines the possibility of quenching of mobility of the incoming projectile. We have demonstrated that appropriate manipulation of a slowness index incorporated in the quasiperiodic off-diagonal modulation may lead to a re-entrant localization-delocalization transition associated with a single-particle mobility edge in the eigenspectrum. The results are corroborated with the standard evaluation of inverse participation ratio, fractal dimension, spectrum, dynamical response and multifractal analysis. The spectrum also reveals the formation of quantum butterfly with respect to the variation of the frequency of the aperiodic overlap integral. The edge states formed in the disallowed regime of the spectrum, are found to be robust against application of on-site disorder. A subtle control over the parameters therefore may offer certain interesting spectral property for this two-port ladder geometry.


\begin{thebibliography}{99}

\bibitem{Anderson1958}
P. W. Anderson,
``Absence of Diffusion in Certain Random Lattices,''
Phys. Rev. \textbf{109}, 1492 (1958).

        \bibitem{Yablonovitch1987}
E. Yablonovitch,
Phys. Rev. Lett. \textbf{58}, 2059 (1987).

\bibitem{John1987}
S. John,
Phys. Rev. Lett. \textbf{58}, 2486 (1987).

\bibitem{Montero1998}
F. R. Montero de Espinosa, E. Jiménez, and M. Torres,
Phys. Rev. Lett. \textbf{80}, 1208 (1998).

\bibitem{Vasseur1998}
J. O. Vasseur, P. A. Deymier, G. Frantziskonis,
G. Hong, B. Djafari-Rouhani, and L. Dobrzynski,
J. Phys.: Condens. Matter \textbf{10}, 6051 (1998).

\bibitem{Tao2007}
A. Tao, P. Sinsermsuksakul, and P. Yang,
Nat. Nanotechnol. \textbf{2}, 435 (2007).

\bibitem{Christ2007}
A. Christ, Y. Ekinci, H. H. Solak, N. A. Gippius,
S. G. Tikhodeev, and O. J. F. Martin,
Phys. Rev. B \textbf{76}, 201405(R) (2007).

\bibitem{Barinov2009}
I. O. Barinov, A. P. Alodzhants, and S. M. Arakelian,
Quantum Electron. \textbf{39}, 685 (2009).

\bibitem{Grochol2008}
M. Grochol and C. Piermarocchi,
Phys. Rev. B \textbf{78}, 035323 (2008).

\bibitem{Damski2003}
B. Damski, J. Zakrzewski, L. Santos, P. Zoller, and
M. Lewenstein,
Phys. Rev. Lett. \textbf{91}, 080403 (2003).

\bibitem{Billy2008}
J. Billy, V. Josse, Z. Zuo, A. Bernard,
B. Hambrecht, P. Lugan, D. Clément,
L. Sanchez-Palencia, P. Bouyer, and A. Aspect,
Nature \textbf{453}, 891 (2008).

\bibitem{Roati2008}
G. Roati, C. D'Errico, L. Fallani, M. Fattori,
C. Fort, M. Zaccanti, G. Modugno,
M. Modugno, and M. Inguscio,
Nature \textbf{453}, 895 (2008).

\bibitem{Aubry1980}
S. Aubry and G. André,
in \textit{Group Theoretical Methods in Physics},
edited by L. Horwitz and Y. Ne'eman,
Annals of the Israel Physical Society, Vol. 3
(American Institute of Physics, New York, 1980), p. 133.


		
		\bibitem{DasSarma1986}
		S. Das Sarma, A. Kobayashi, and R. E. Prange,
		Phys. Rev. Lett. \textbf{56}, 1280 (1986).
		
		\bibitem{Biddle2010}
		J. Biddle and S. Das Sarma,
		Phys. Rev. Lett. \textbf{104}, 070601 (2010).
		
		\bibitem{Ganeshan2015}
		S. Ganeshan, J. H. Pixley, and S. Das Sarma,
		Phys. Rev. Lett. \textbf{114}, 146601 (2015).
		
		\bibitem{Sun2015}
		M. L. Sun, G. Wang, N. B. Li, and T. Nakayama,
		Europhys. Lett. \textbf{110}, 57003 (2015).
		
		\bibitem{Gopalakrishnan2017}
		S. Gopalakrishnan,
		Phys. Rev. B \textbf{96}, 054202 (2017).
		
		\bibitem{Purkayastha2017}
		A. Purkayastha, A. Dhar, and M. Kulkarni,
		Phys. Rev. B \textbf{96}, 180204(R) (2017).
		
		\bibitem{Boers2007}
		D. J. Boers, B. Goedeke, D. Hinrichs, and M. Holthaus,
		Phys. Rev. A \textbf{75}, 063404 (2007).
		
		\bibitem{Li2017}
		X. Li, X. Li, and S. Das Sarma,
		Phys. Rev. B \textbf{96}, 085119 (2017).
		
		\bibitem{Luschen2018}
		H. P. Lüschen, S. Scherg, T. Kohlert, M. Schreiber, P. Bordia,
		X. Li, S. Das Sarma, and I. Bloch,
		Phys. Rev. Lett. \textbf{120}, 160404 (2018).
		
		\bibitem{An2018}
		F. A. An, E. J. Meier, and B. Gadway,
		Phys. Rev. X \textbf{8}, 031045 (2018).
		
		\bibitem{An2021}
		F. A. An, K. Padavić, E. J. Meier, S. Hegde, S. Ganeshan,
		J. H. Pixley, S. Vishveshwara, and B. Gadway,
		Phys. Rev. Lett. \textbf{126}, 040603 (2021).
		
		\bibitem{Roy2021}
		S. Roy, T. Mishra, B. Tanatar, and S. Basu, Phys. Rev. Lett. \textbf{126}, 106803 (2021).
		
\bibitem{Macia2006}
E. Maciá,
Phys. Rev. B \textbf{74}, 245105 (2006).

\bibitem{Cuniberti2007}
G. Cuniberti, E. Maciá, A. Rodriguez, and R. A. Römer,
arXiv:0707.3224.

\bibitem{Mrevlishvili1998}
G. M. Mrevlishvili,
Thermochim. Acta \textbf{308}, 49 (1998).

\bibitem{Moreira2006}
D. A. Moreira, E. L. Albuquerque, and C. G. Bezerra,
Eur. Phys. J. B \textbf{54}, 393 (2006).
	
	
\bibitem{Goblot2020}
V. Goblot, A. Štrkalj, N. Pernet, J. L. Lado, C. Dorow,
A. Lemaître, L. Le Gratiet, A. Harouri, I. Sagnes,
S. Ravets, A. Amo, J. Bloch, and O. Zilberberg,
Nat. Phys. \textbf{16}, 832 (2020).
\bibitem{Li2020}
X. Li and S. Das Sarma,
Phys. Rev. B \textbf{101}, 064203 (2020).
\bibitem{Katsanos1995}
D. E. Katsanos, S. N. Evangelou, and S. J. Xiong,
``Quantum electron dynamics in periodic and aperiodic sequences,''
Phys. Rev. B \textbf{51}, 895 (1995).
\end{thebibliography}
\end{document}